\documentstyle[11pt,paspconf,epsf]{article}
\epsfclipon
\begin{document}

\title{Evolutionary behaviour of BL Lac objects}
\author{V. Beckmann}
\affil{Hamburger Sternwarte, Gojenbergsweg 112, 21029 Hamburg, Germany}

\begin{abstract}
We present analysis of a new flux limited sample of 72 \mbox{X-ray} selected 
BL Lacertae objects with known redshifts for 59 of them. 
We find the more X-ray dominated BL Lac having negative evolution with redshift, while 
intermediate objects show no evolution at all. \mbox{X-ray} dominated objects show 
higher X-ray luminosities and flatter X-ray spectra than the intermediate ones. We were also able to
determine the redshift of 1517+656, which is part of our sample and is the
optically most luminous X-ray selected BL Lac known so far. 
\end{abstract}


\section{Introduction}

BL Lac objects are a rare type of Active Galactic Nuclei (AGN). Because their spectrum does not show
prominent emission and absorption lines they are difficult to find and getting
redshift information from absorption lines due to their host galaxies requires
a huge amount of telescope time. Therefore not many complete samples exist with known
redshifts. The most
often cited ``$1 \; {\rm Jy}$ Sample'' contains 35 BL Lac objects selected in the
radio wavelength region (Stickel et al.\ 1991). At X-ray wavelengths a complete
well defined sample was built up with the EMSS (Stocke et al.\, 1991); Morris
et al.\ (1991) presented a sample of 22 objects down to $f_{\rm X}(0.3 - 3.5 \; {\rm keV}) = 5 \cdot 10^{-13} \; {\rm erg} \; {\rm cm}^{-2} \; {\rm sec}^{-1}$
which was later expanded to 30 objects and lower fluxes by Wolter et
al.\ (1994). While radio selected BL Lac show positive evolution, like ``normal'' AGN, X-ray selected BL Lac show negative evolution (Morris et
al.\ 1991, Wolter et al.\ 1994), so 
they have been less numerous and/or luminous at high redshifts than in
the local universe. 
We contribute to the discussion about the evolutionary behaviour of BL Lac objects with a new sample of up
to now 72 X-ray selected BL Lac objects, with redshifts available for more
than 80 \% of them. The selection is based on the ROSAT All-Sky Survey (RASS,
Voges et al.\ 1996).\\
In this paper cosmological parameters $H_{0} = 50\, {\rm km\, sec}^{-1} \,
{\rm Mpc}^{-1}$ and $q_{0} = 0.5$ are used.

\section{The HRX BL Lac sample}

\begin{figure}
\plotone{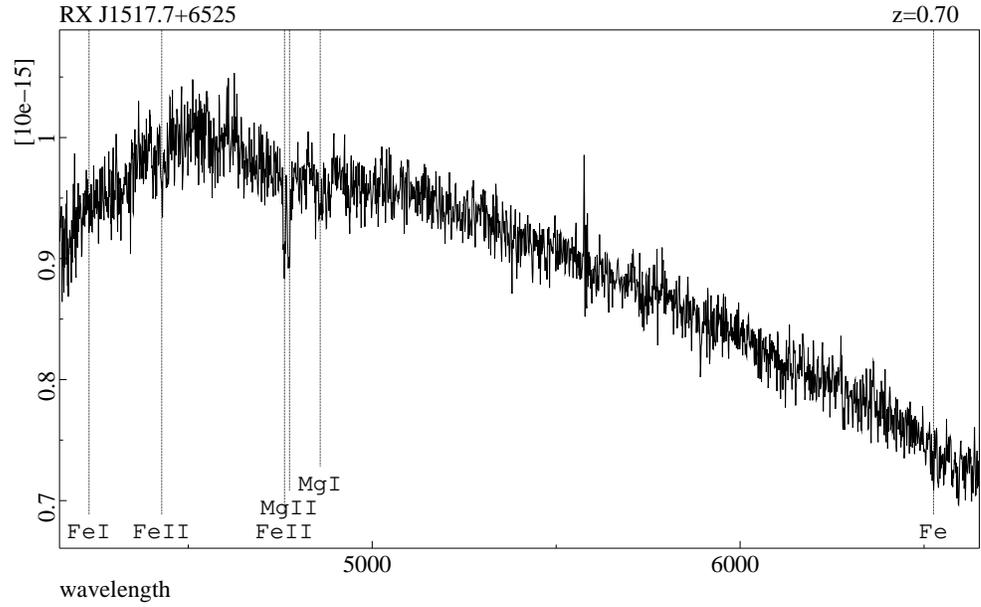}
\caption{The spectrum of 1517+656, taken with the Calar Alto 3.5m telescope
  and MOSCA focal reducer.} \label{fig-1}
\end{figure}

The BL Lac sample presented here is part of the Hamburg-RASS X-ray Bright AGN
sample (HRX, Cordis et al.\ 1996). The HRX
was pre-identified on direct and objective prism plates which were taken with
the Hamburg Schmidt telescope on Calar Alto, Spain, within the Hamburg Quasar
Survey (HQS, Hagen et al.\ 1995). Several follow-up campaigns were done to
classify those objects which had been unidentified on the Schmidt plates and
to determine redshifts for the AGN candidates. 
A first BL Lac sample, hereafter core sample, drawn from an area of $1687 \;
{\rm deg}^{2}$ contained 35 objects and was complete down to a
ROSAT PSPC ($0.5 - 2$ keV) countrate of $0.075 \; {\rm cts} \; {\rm
  sec}^{-1}$. This sample was presented and discussed by Bade et al.\ (1998).
To examine the dependence of the evolutionary behaviour on X-ray
dominance, we split the core sample into two groups according to
$\alpha_{\rm ox}$\footnote{we define $\alpha_{\rm ox}$ as the power law index between
  $1$ keV and $4400$ {\AA} with $f_{\nu} \propto \nu^{-\alpha_{\rm ox}}$}. Objects
with $\alpha_{\rm ox} < 0.9$ we will call X-ray dominated BL Lac, the other
ones are the {\em intermediate objects} (IBL), because they have physical
properties between the X-ray and radio
dominated BL Lac objects. We found the X-ray dominated objects to have
negative evolution, while objects with steeper $\alpha_{\rm ox}$ showed
no evolution at all. 
A new sample, hereafter enlarged sample, has been defined now to improve the
statistics. It contains the core sample and 37 further BL Lac drawn from an
area of $\sim 2800 \; {\rm deg}^{2}$ with countrate limit of $0.09 \; {\rm
  cts} \; {\rm sec}^{-1}$. Spectroscopic observations to determine their
redshifts are underway. Prensently for the core region redshifts are known for
32 out of 35 BL Lac (91 \%) and for the enlarged region for 59 out of 72 BL
Lac (82 \%). 
Our criteria to classify X-ray sources as BL Lac objects are similar to the 
work of the EMSS. So we use ``conservative'' criteria: emission lines with 
\mbox{$W_{\lambda} > 5$ {\AA}} must be absent and the contrast of the Ca II 
break from the hosting galaxy must be less than $25 \%$ in the optical 
spectrum. 
We want to emphasize that the core sample was selected without help of
information on radio fluxes. It was found however that at the flux limit of the
HRX sample all BL Lac objects were detected by the VLA surveys NVSS (Condon et
al.\ 1998) and FIRST (White et al.\ 1997). We therefore used these radio surveys to find the BL Lac objects in the enlarged area. After completion of the HRX
also the enlarged sample will be determined without the help of the radio
information. 
The X-ray selection is based on the ROSAT
PSPC hard countrate in the RASS. The conversion factor between flux and
countrate for a single power law depends on the photon
index\footnote{the energy index $\alpha_{\rm E}$ is related to the photon index $\Gamma
  = \alpha_{\rm E}+ 1$} $\Gamma$ and the
hydrogen column density $N_{\rm H}$. We assume only absorption due to Galactic
$N_{\rm H}$ taken from Stark et al.\ (1992). The photon index was calculated with
the hardness ratios given in the RASS; this method is described in detail in Schartel et al.\ (1996).\\     
Spectroscopic follow-up observations of BL Lac are made with the MOSCA
spectrograph at the Calar Alto 3.5m telescope. A first result is the
determination of the redshift of the BL Lac 1517+656 (Fig.~\ref{fig-1}).
\begin{figure}[!tbp]
\plotone{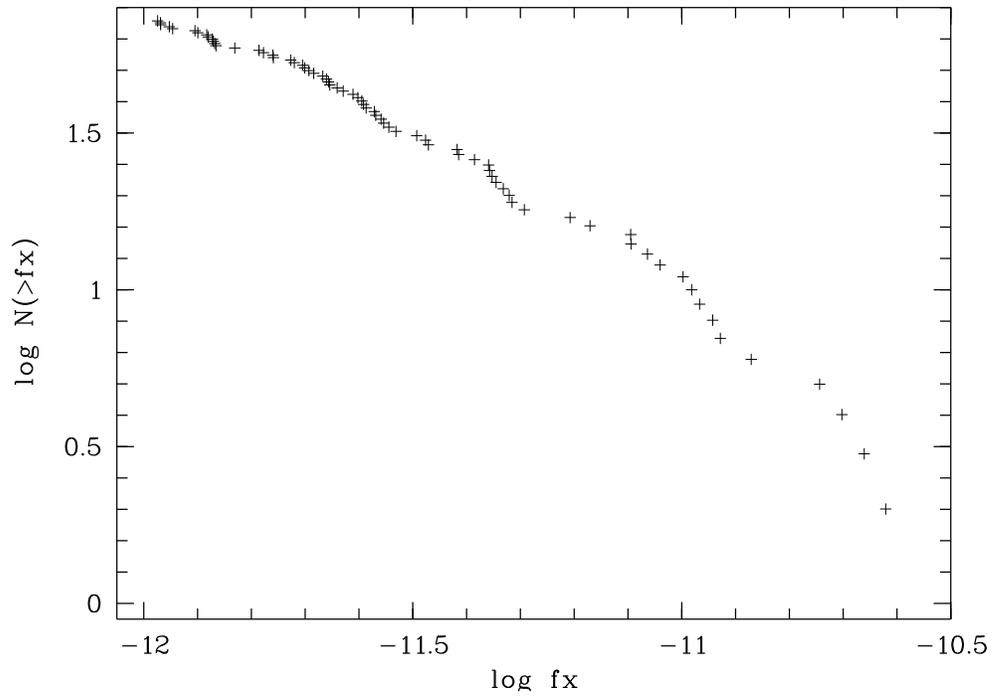}
\caption{$\log N(>f_{\rm X}) \mbox{---} \log f_{\rm X}$ distribution for the 72 BL Lac objects of the
  enlarged HRX BL Lac sample.} \label{fig-2}
\end{figure}

\section{Analysis}

The $\log N(>f_{\rm X}) \mbox{---} \log f_{\rm X}$ distribution (Fig.~\ref{fig-2}) is independent of
the subtle redshift determination process for BL Lac objects. The sample shows
a steep slope consistent with the Euclidean value of $-1.5$ down to
$f_{\rm X}$(0.5--2.0 keV) $\simeq 8 \cdot 10^{-12} \; {\rm erg} \; {\rm cm}^{-2} \; {\rm sec}^{-1}$. Below this
flux the distribution has a flatter slope of $-1.0$. The bump around the turnover
point is mainly caused by X-ray dominated objects. This can be clearly seen by
deriving number counts for the two subgroups, related to the X-ray dominance as
shown in Figure 3. The IBL show a more or less flat
slope of $-1.0$ in the range
$-12 \le \log{f_{\rm X}} \le -10.5$, while the X-ray dominated objects
show a steep slope of -2.1 for $f_{\rm X} \ge 8 \cdot 10^{-12}
\; {\rm erg} \; {\rm cm}^{-2} \; {\rm sec}^{-1}$, and a flat slope ($-0.7$) for
lower fluxes. The different distributions indicate that the evolution of the
two BL Lac groups is different.\\
\begin{figure}
\plottwo{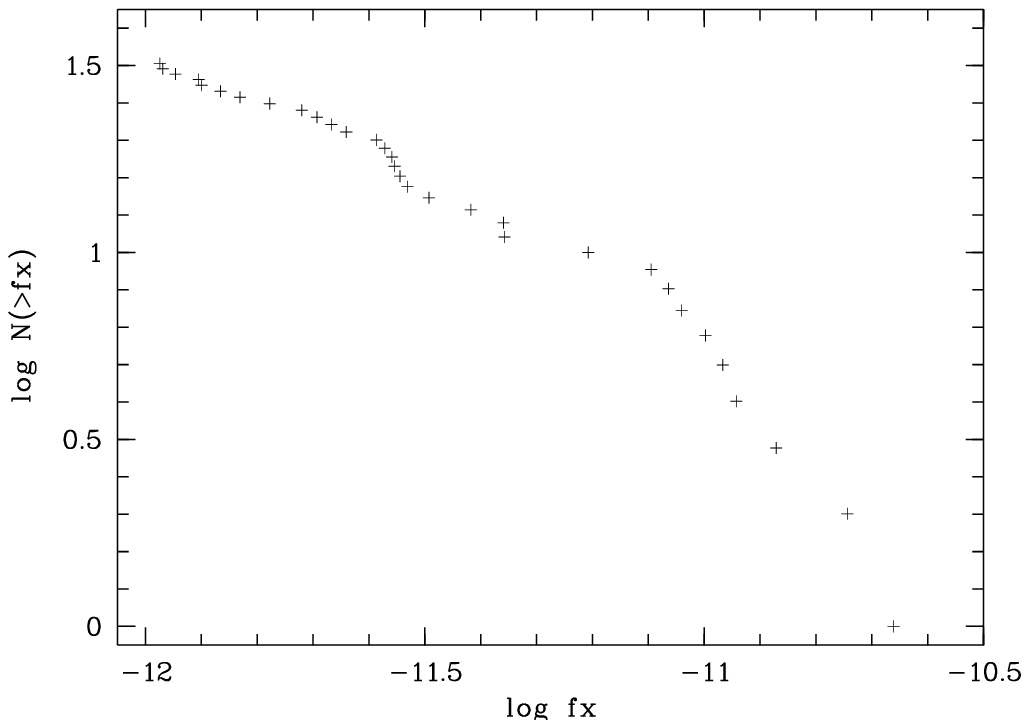}{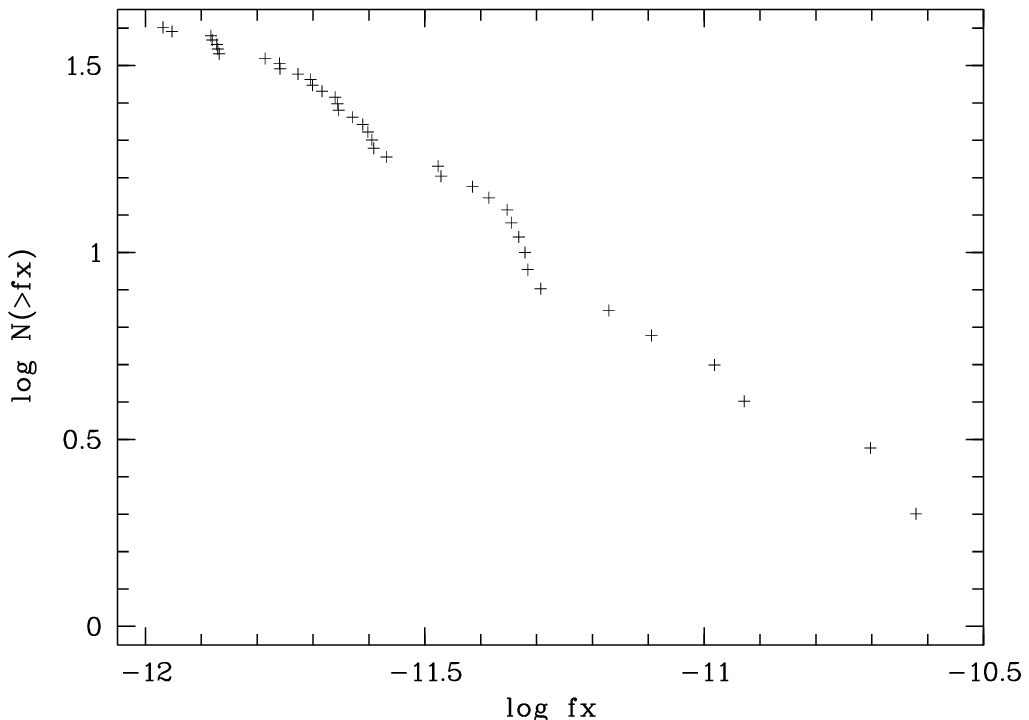}
\caption{Number counts for X-ray dominated objects ($\alpha_{\rm ox} < 0.9$, left
  panel) and intermediate objects ($\alpha_{\rm ox} > 0.9$, right
  panel).} \label{fig-3}
\end{figure}
While we have reliable redshifts for all
but one of the X-ray dominated objects, we still miss redshifts for 12
IBL; in the view of the unified scheme due to different
opening angles of X-ray and radio selected BL Lac, the reason might be, that
in the IBL the core outshines the hosting galaxy and
absorption lines are more difficult to detect than in the X-ray dominated BL
Lac which are observed more off-axis from the jet.\\
To examine the evolutionary behaviour we use a $\langle V/V_{\rm max} \rangle$ test (Schmidt 1968)
which is the mean of the ratios between the volume, in which an object is found, and
the maximal volume, where the object could have been found. This depends
mainly on the ratio between the observed flux and the flux limit. The results
are listed below:\\

\begin{tabular}{l|c|c}
$\langle V/V_{\rm max} \rangle$ & Intermediate & X-ray dominated\\
 & $\alpha_{\rm ox} > 0.9$ & $\alpha_{\rm ox} < 0.9$\\
\hline
core sample & $0.48 \pm 0.08$ & $0.34 \pm 0.06$ \\
enlarged sample & ? & $0.30 \pm 0.04$ \\
\end{tabular} \\

The $\langle V/V_{\rm max} \rangle$ test confirms the negative
evolution for the enlarged sample reported by Bade et al.\ (1998). For the IBL
redshifts are missing for 30\% of the objects so that their non-evolution 
could not be tested with the enlarged sample yet.
\begin{figure}
\plottwo{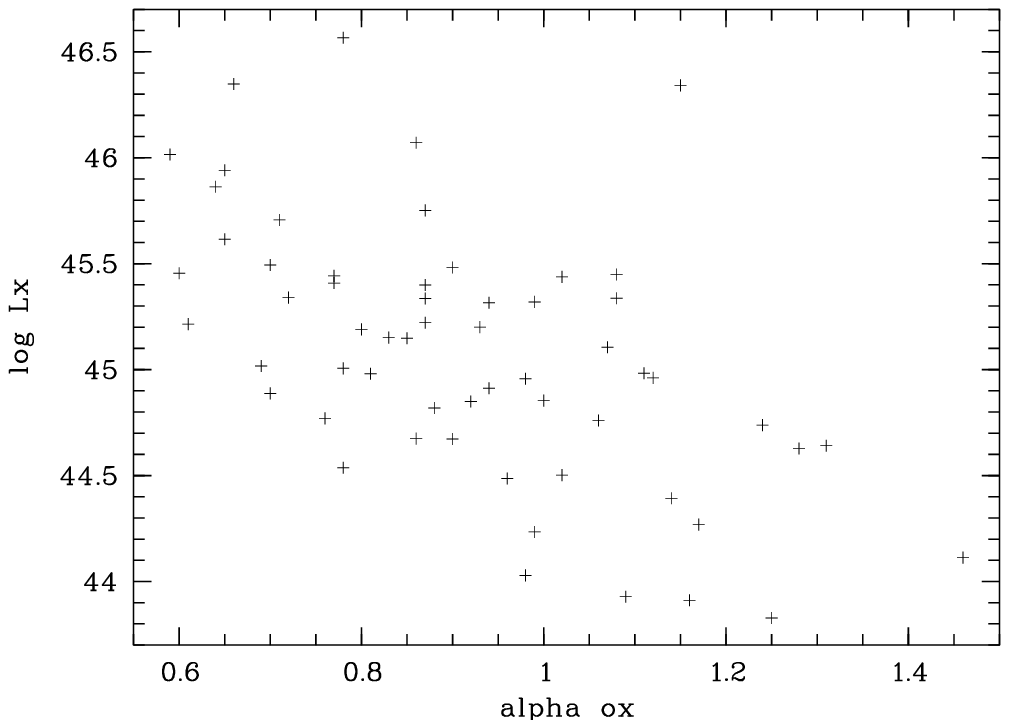}{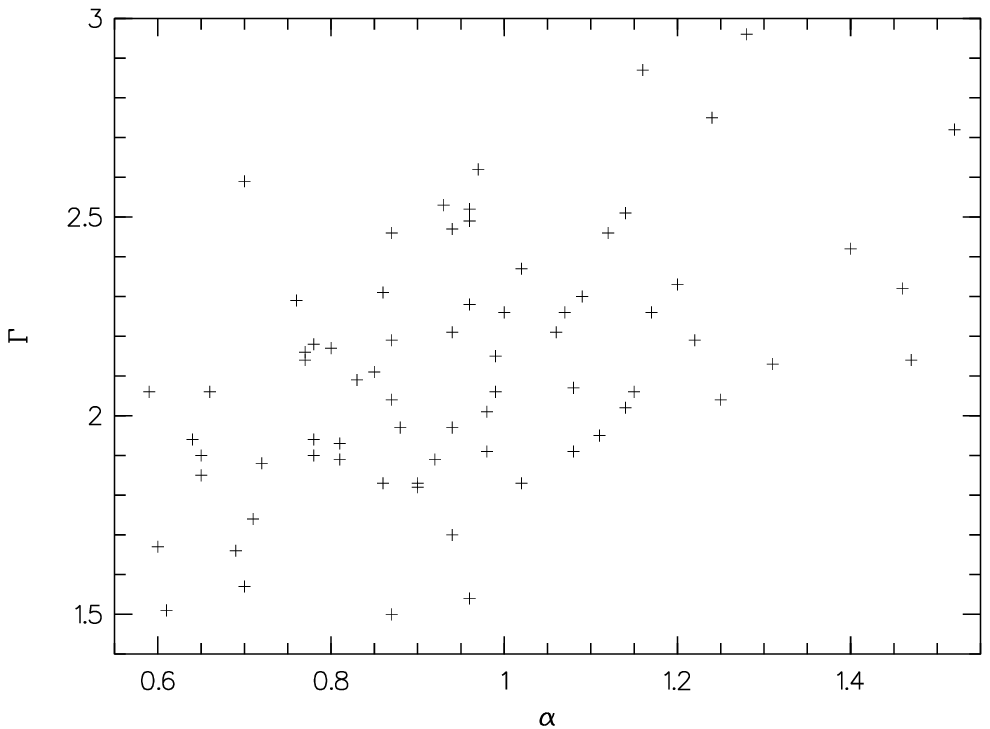}
\caption{X-ray luminosity in the hard ROSAT PSPC band (left panel) and
  the X-ray spectral slope (right panel) versus $\alpha_{\rm ox}$. X-ray dominated
  BL Lac have higher X-ray luminosities than IBL and their X-ray spectra are
  flatter.} \label{fig-4}
\end{figure}
The X-ray dominated BL Lac have in the mean higher redshifts, which implies
higher luminosities as well. Figure~\ref{fig-4} shows the dependence of the X-ray luminosity on
$\alpha_{\rm ox}$. There is clear evidence for a relation between those values on
a $> 99\%$ confidence level, while the absolute optical magnitude and the luminosity in the radio band seem not
to be correlated with $\alpha_{\rm ox}$ (Bade et al.\ 1998).\\
The spectral properties in the ROSAT PSPC band are quite typical for 
\mbox{X-ray} selected BL Lac; we find a photon index $\langle \Gamma \rangle =
1.97 \pm 0.26$ for the X-ray dominated objects and  $\langle \Gamma \rangle = 
2.24 \pm 0.31$ for IBL (Fig.~\ref{fig-4}), which is in good agreement with 
Siebert et al.\ (1998), who found a value of
$\langle \Gamma \rangle = 2.35 \pm 0.55$ for the latter. These spectral slopes
are slightly
flatter than the values found for all types of X-ray selected AGN
(e.g. $\langle \Gamma \rangle = 2.50 \pm 0.48$, Walter \& Fink 1993; $\langle
\Gamma \rangle = 2.42 \pm 0.44$, Ciliegi \& Maccacaro 1996; $\langle \Gamma \rangle = 2.53 \pm 0.42$, Beckmann \& Bade 1999).\\
While determining redshifts for objects of our sample, we also took an optical
spectrum of 1517+656. This source was identified as a BL Lac in the course of the Einstein Slew Survey (Elvis et al.\ 1992). We determined the
redshift as $z = 0.702$ (Fig.~\ref{fig-1}) and the apparent
magnitude as $V = 15.9 \; {\rm mag}$ by comparison with standard stars near to
the BL Lac (Villata et al.\ 1998). With an absolute magnitude of
$M_{V} =
-27.5$ this is the optically most luminous X-ray selected BL Lac which is known so far.
 
\section{Discussion}

For the X-ray dominated BL Lac we find negative evolution, while IBL show no 
evolution at all, even though the latter result has still to be
confirmed for the enlarged sample. Radio selected BL Lac show
positive evolution; therefore it is hard to imagine how a unified scheme for
BL Lac could look like, which tries to identify X-ray and
radio selected BL Lac objects with a common parent population in which
only some physical parameters of the jet are varied (e.g. beaming properties, angle
$\Theta_{C}$ between line of sight and jet). The different evolutionary
behaviour is a strong hint 
to physically different types of objects. The distinct properties in the two
X-ray selected subgroups are already revealed in the bright part of the $\log N(>f_{\rm X})
\mbox{---} \log f_{\rm X}$
distribution. It is therefore implausible that selection effects, as described
by Browne \& Marcha (1993), are responsible for these differences. It might be,
that the evolving time for X-ray bright jets is longer than for the radio
fraction; this would explain, why X-ray selected BL Lac are more numerous in
the local universe than at high redshift. 
On the other hand the relation between
$\alpha_{\rm ox}$ and the X-ray luminosity leads to the idea of an outburst
scenario. In that case, the intermediate objects would be the parent
population and BL Lac in an X-ray outburst would have high X-ray luminosities
and would be X-ray dominated. Strong evidence for this model is
supplied by the fact
that there are no X-ray dominated BL Lac with low X-ray luminosities; because
we only miss the redshift of one of those objects, there is little doubt about
this result. In this model, an IBL having an outburst would
become more luminous in the X-rays, while staying nearly constant in the
optical and in the radio band; therefore the $\alpha_{\rm ox}$ value would
decrease and the spectral slope in the X-ray band would flatten.
This model is also supported by the spectral
slope distribution, showing the outbursting BL Lac having flatter X-ray
spectra than the intermediate ones (Fig.~\ref{fig-4}), while the absolute optical magnitude seems
not to be correlated with $\alpha_{\rm ox}$.\\
Both scenarios as well as the unified scheme still do not give satisfying
explanation for all observed properties in BL Lac objects.

\acknowledgements

First of all I would like to thank Norbert Bade, who started this
project. Also thanks to S. Laurent-Muehleisen and A. Schwope for redshift
information and to H.-J. Hagen for providing the optical reduction software.

\end{document}